\documentclass[USletter,conference]{IEEEtran}

\IEEEoverridecommandlockouts
\usepackage{mathrsfs}
\usepackage{cite}
\usepackage{fancyhdr}
\usepackage{geometry}
\usepackage{amsmath,amssymb,amsfonts}
\usepackage{algpseudocode,algorithm}
\usepackage{graphicx}
\usepackage{textcomp}
\usepackage{xcolor}
\usepackage{subcaption}
\usepackage{array}

\usepackage{booktabs}

\def\BibTeX{{\rm B\kern-.05em{\sc i\kern-.025em b}\kern-.08em
    \kern-.1667em\lower.7ex\hbox{E}\kern-.125emX}}
\begin{document}
\title{Bone Feature Segmentation in Ultrasound Spine Image with Robustness to Speckle and Regular Occlusion Noise\\
}
\author{\IEEEauthorblockN{Zixun Huang$^1$\thanks{$^1$Z. Huang, L.W. Wang, F. H. F. Leung are with the Department of Electronic and Information Engineering, The Hong Kong Polytechnic University, Hong Kong, China  zixun.huang@connect.polyu.hk, liwen1.wang@connect.polyu.hk, frank-h-f.leung@polyu.edu.hk}, Li-Wen Wang$^1$, Frank H. F. Leung$^1$, 
		Sunetra Banerjee$^2$\thanks{$^2$S. Banerjee, S.H. Ling are with the School of Biomedical Engineering, University of Technology Sydney, NSW, Australia  sunetra.banerjee@student.uts.edu.au, steve.ling@uts.edu.au}, 
		De Yang$^3$\thanks{$^3$D. Yang, T. Lee, Y.P. Zheng are with the Department of Biomedical Engineering, The Hong Kong Polytechnic University, Hong Kong, China  de.de.yang@connect.polyu.hk, timothy.lee@connect.polyu.hk, yongping.zheng@polyu.edu.hk}, Timothy Lee$^3$, \\
		Juan Lyu$^4$\thanks{$^4$J. Lyu is with the College of Information and Communication Engineering, Harbin Engineering University, Harbin, 150001, China lvjuan@hrbeu.edu.cn}, 
		Sai Ho Ling$^2$, Yong-Ping Zheng$^3$}}

\maketitle
\begin{abstract}

3D ultrasound imaging shows great promise for scoliosis diagnosis thanks to its low-costing, radiation-free and real-time characteristics. The key to accessing scoliosis by ultrasound imaging is to accurately segment the bone area and measure the scoliosis degree based on the symmetry of the bone features. The ultrasound images tend to contain many speckles and regular occlusion noise which is difficult, tedious and time-consuming for experts to find out the bony feature. In this paper, we propose a robust bone feature segmentation method based on the U-net structure for ultrasound spine Volume Projection Imaging (VPI) images. The proposed segmentation method introduces a total variance loss to reduce the sensitivity of the model to small-scale and regular occlusion noise. The proposed approach improves 2.3\% of Dice score and 1\% of AUC score as compared with the u-net model and shows high robustness to speckle and regular occlusion noise.

\end{abstract}
\providecommand{\Keywords}[1]{\textbf{{\textit{Keyword--}}}#1}

\section{Introduction}
Scoliosis is a condition in which the spinal cord gets severely deformed over time. The process of detection and diagnosis of the condition has been around for a long time\cite{r28}. A detection process involves scanning the spinal area of a patient with a suitable modality and accessing the curvature of the spine. This process is repeated over time and if, at any stage, the curvature of spine is detected to exceed 10°, the patient is categorized for potential scoliosis treatment.

The highest risk factor for scoliosis occurs in teenagers since their skeletal structure is still under development. This condition in medical parlance is known as Adolescent Idiopathic Scoliosis (AIS) \cite{r1}. If left untreated, AIS not only has a physiological impact on the adolescent body structure (such as uneven shoulders and misalignment of hips) but also has severe psychological impact and hindrance to the well-being of the adolescent patient \cite{r14}. 

 The current practice of detection and diagnosis of AIS generally involves (a) repeated scanning using X-Rays and (b) measuring the Cobb Angle, which is a gold standard to assess and monitor AIS \cite{r2}. This practice however leads to repeated exposure to radiation that is especially harmful to adolescents. During the whole course of scoliosis treatment, they may be exposed up to 25 X-rays sessions\cite{r3}. In fact, Levy et al. \cite{r4} demonstrated that AIS patients who had undergone multiple X-rays have a 2.4\% higher risk of developing cancer than normal young people.

Other modalities such as MRI are also found unsuitable as they are time-consuming, expensive and cannot meet the needs of large-scale screening and frequent assessments\cite{r5}. Alternatively, to detect the bony feature of adolescent teenagers, non-radiation imaging techniques, such as ultrasound imaging, are being used. Though the noise in the output images is higher, such techniques are not only safe but also affordable and fast. 

\begin{figure*}[t]
	
	\centerline{\includegraphics[keepaspectratio, width=1\textwidth]{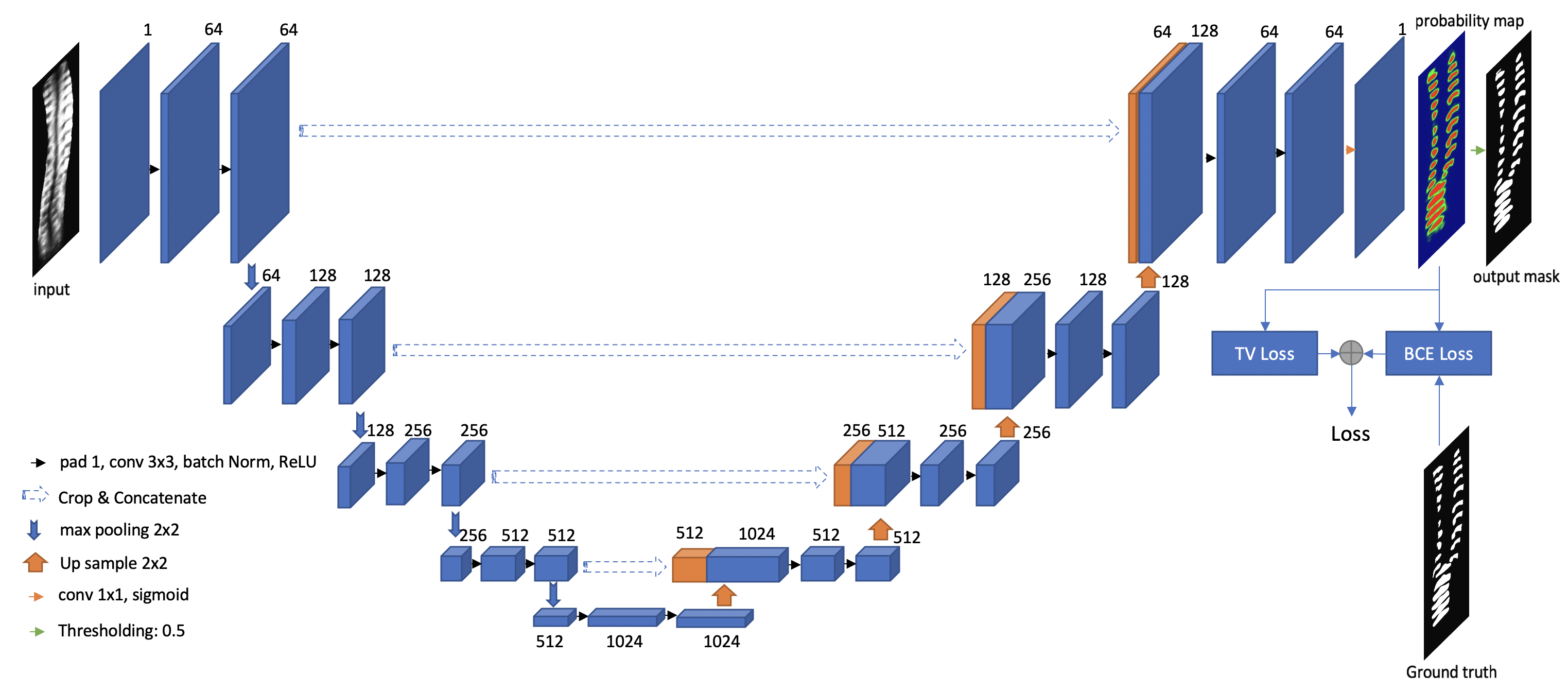}}
	\caption{The overall architecture of the proposed bone feature segmentation model of RSN-U-net.}
	\label{f1}
\end{figure*}

We pioneered the technique of 3D ultrasound imaging to assess scoliosis and have achieved great success\cite{sco_cheung2015us, sco_zheng2016relia, sco_zhou2015, sco_zhou2017auto, r21}. The scolioscan system is a radiation-free, semi-automatic 3D ultrasound system. The imaging technique used in scolioscan is volume projection imaging (VPI). VPI analyses the intensity of all voxels of ultrasound volumetric data to form a coronal image \cite{sco_cheung2015us}. The spinal angle measured by scolioscan using VPI has been proven comparable to the gold standard of Cobb angle obtained through the X-ray method. 

While this has been ground-breaking research, the scolioscan system requires manual measurement of the spine angle, which in turn depends on the judgment and expertise of the doctor or examiner. In fact, it is observed that the variation of the spinal angle measured by different examiners can be as high as 2 - 3°. So, as an improvement, several automatic measurements techniques were developed\cite{sco_zhou2015,sco_zhou2017auto}. In \cite{sco_zhou2015}, the spine curvature angle is obtained by deriving the inflection points.

In \cite{r22,r21}, it has been suggested that the Transverse process (TP) measurement method can be used to measure spinal deformation with 3D ultrasound images. The methodology of TP measurement is to detect the bony features in an ultrasound scan. However, the ultrasound images tend to contain many speckles\cite{r23} and noise (Fig. \ref{f2}) which makes it difficult for experts to demarcate the bony features.



In this paper, a new segmentation method called U-net with robustness to speckle and regular occlusion noise (RSN-U-net) is introduced to effectively segment the bony features in an ultrasound spine image. A new technique called total variance (TV) loss is presented. Through experiments and studies, the TV loss technique is found adept in training the neural network and improving the robustness against the speckle and regular occlusion noise. Through visual comparison of the results, the proposed method is found to achieve a stable and better performance even for cases where the input ultrasound image has high speckles and noise.


This paper is organized into three main sections. In Section II, the proposed RSN-U-net architecture and methodology are discussed. In Section III, an account of the experimental setups along with the dataset and the results are given. Finally, in Section IV, a conclusion is drawn with a discussion on robustness of the model and future work.

\section{Methods}
Recently, U-net has been widely used in medical image segmentation tasks owing to its superior performance\cite{r24}. In this paper, Our bone feature segmentation model is mainly drawn from the U-net structure.
Fig. \ref{f1} shows the flowchart of the proposed segmentation model. Given an ultrasound spine image, our segmentation model aim to estimate a probability map of the bone feature. Then a thresholding strategy is used to generate the segmentation mask of the ultrasound spine image. 

\subsection{Bone Feature Segmentation Model}

Bone feature segmentation aims to find out all the regions of the bone from the original ultrasound image, and the surface of the bone is not fixed. It means that the segmentation method needs to distinguish whether a pixel belongs to a bone or not. It is a pixel-level binary classification task. The output of the model should be a segmentation mask with the same size as the input image. 

As shown in Fig. \ref{f1}, the proposed RSN-U-net adopts the auto-encoder structure. The encoder part consists four repeated encoder stacks. Every encoder stack contains two convolutional layers with the kernel size of $3 \times 3$ and a max pooling layer with the stride of 2. The convolutional layers aim to extract useful bone features from the original gray scale ultrasound image. The down-sampling strategy (The pooling layers) is utilized to select meaningful features and reduce the scale of the feature map which not only can reduce the computation complexity but also give the layers large receptive fields. Considering the down-sampling strategy may lose local information\cite{r31,r32}, the shallow features are cropped and concatenated to the corresponding deep layers to remedy the lost information. The decoder part consists of four repeated decoder stacks. Each decoder stack contains two convolutional layers with the kernel size of $3 \times 3$ and an up-sampling layer with the stride of 2. The convolutional layer are designed to extract bone features from the concatenated features. The up-sampling layers gradually enlarge the features back to the original size. In the last layer, a 1 $\times$ 1 convolutional layer with sigmoid function is used to map and normalize the bone features to segmentation probability map. Finally, the segmentation probability map is obtained through a binary operation with the threshold of 0.5.

\begin{figure}[t]
	
	\centerline{\includegraphics[width=0.25\textheight, height=0.35\textwidth]{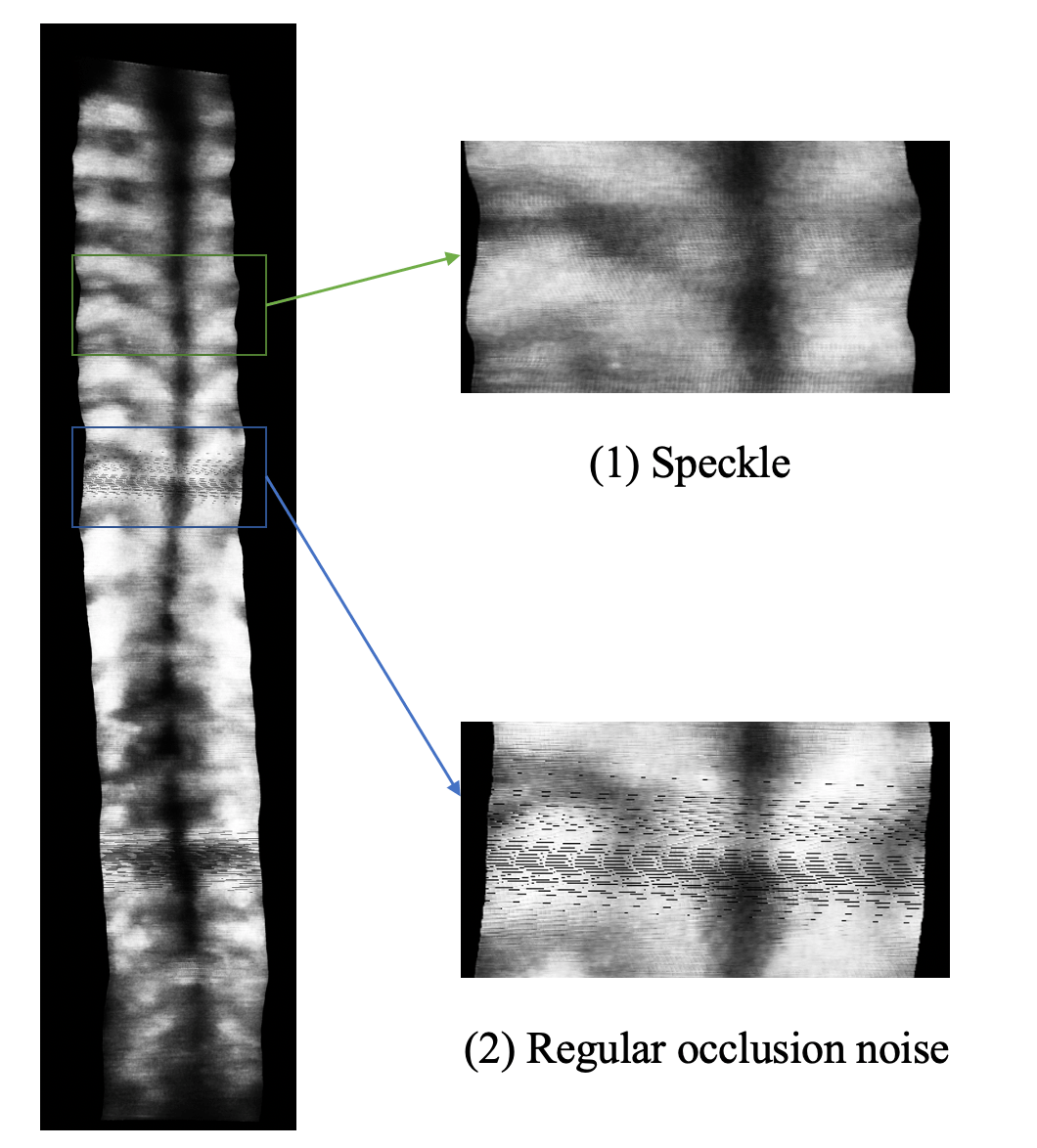}} 
	\caption{Example of the speckles and regular occlusion noise of the ultrasound spine image.}
	\label{f2}
\end{figure}

\begin{figure*}[htp]
	\centerline{\includegraphics[ width=1\textwidth, height=0.55\textwidth]{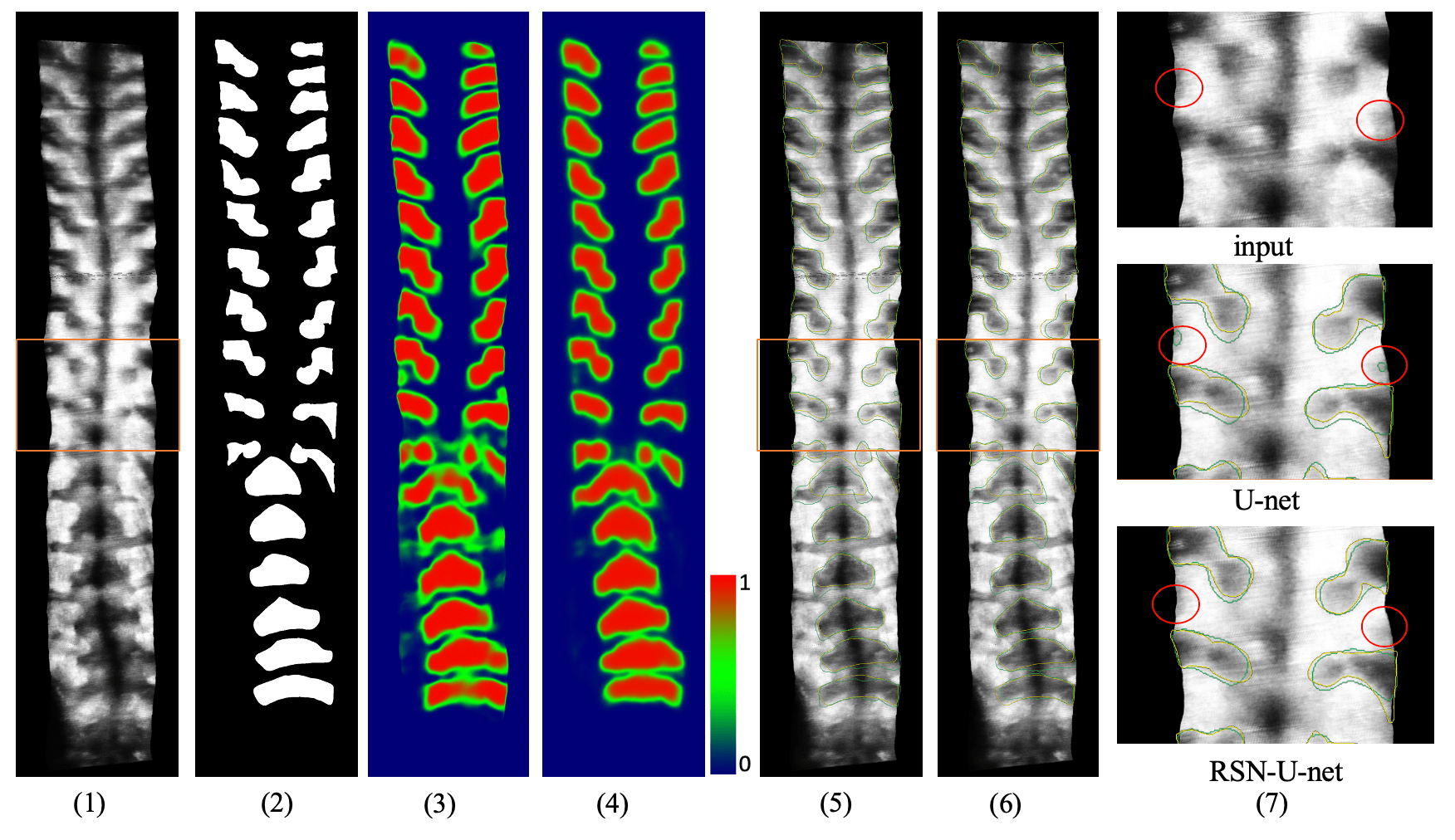}} 
	\caption{Visual comparison of U-net and RSN-U-net for bone feature segmentation.(1) the original input VPI image; (2) the ground truth mask; (3) the foreground probability map of the prediction result by U-net (visualized by heatmap); (4) the foreground probability map of the prediction result by RSN-U-net; (5) segmentation result by U-net (Dice score: 0.7827); (6) segmentation result by RSN-U-net (Dice score: 0.8144); (7) comparison of the segmentation results on a part of ultrasound image with speckles (up to down: the input image, the segmentation result of U-net, the segmentation result of RSN-U-net) (Green line: segmentation result of the model; Yellow line: Ground Truth) (zoom in for better view).}
	\label{f3}
\end{figure*}

\subsection{Loss Function}

The original ultrasound VPI image is a 2-D mapping image extracted from 3-D ultrasound voxels, which contain 2000-2500 2-D ultrasound images. The VPI images used for segmentation usually contain regular occlusion noise which may be caused by projection as shown in Fig. \ref{f2} (2). Furthermore, the ultrasonic image tends to have many speckles which are shown in Fig. \ref{f2} (1). Both regular occlusion noise and speckles will reduce the performance of the segmentation model. Therefore, in order to enhance the robustness of the model to the noise and speckle, based on the traditional binary cross-entropy (BCE) loss function($\mathcal{L}_{B}$),  we introduce Total Variance (TV) Loss function($\mathcal{L}_{TV}$) to constrain the estimation of the model. The loss function $(\mathcal{L})$  is described in the right part of Fig. \ref{f1}. Mathematically, it can be written as: 
 
\begin{equation}
\mathcal{L} (\textbf{Y},\hat{\mathbf{Y}})=\mathcal{L}_{B}(\textbf{Y},\hat{\mathbf{Y}}) + \lambda \mathcal{L}_{TV}(\hat{\mathbf{Y}})
\end{equation}

\noindent where $\hat{\mathbf{Y}}\in \mathbb{R}^{W\times H}$ represents the estimated segmentation probability map that is produced by the proposed RSN-U-net. The \textit{H} and \textit{W} mean the height and width of the estimated segmentation probability map. The term $\textbf{Y} \in \mathbb{R}^{W\times H}$ represents the ground truth segmentation mask. A weight parameter $\lambda$ is utilized to balance the smoothness constraint. (Empirically, $\lambda$ is set to 0.4 in our experiments)

\subsubsection{Pixel-level Classification}
The segmentation network is designed to find the area that belongs to the bone. In order to enhance the classification ability of the network for each pixel, we use the Binary Cross-Entropy (BCE) loss function to calculate the classification error of each pixel. The value of binary cross-entropy loss $\mathcal{L}_{B}$ is the negative average of the error of every pixel in the output probability map, viz.

\begin{small}
\begin{equation}
\mathcal{L}_{B}(\textbf{Y},\hat{\mathbf{Y}}) =-\frac{\sum_{i=1}^{H}\sum_{j=1}^{W}(y_{i,j} log(\hat{y}_{i,j})+(1-y_{i,j})  log(1-\hat{y}_{i,j}))}{H\times W}
\end{equation}
\end{small}

\noindent where $H \times W$ is the total number of pixels in the original ultrasound spine image. $\hat{y}_{i,j}$ and $y_{i,j}$ are the estimated probability and the ground truth respectively to the position in (\textit{i}, \textit{j}).  

\subsubsection{Robustness to Speckle and Noise}
The regular occlusion noise and speckles in the original ultrasound spine image will cause the segmentation network to erroneously classify the noisy area as a bone feature area, thereby generating a misclassified area that is much smaller relative to the bone feature area. In other words, since the pixel values in speckle and regular occlusion noise area vary frequently, incorrectly classified regions tend to have a higher proportion of edge pixels with high-frequency energy. The total variance loss calculates the total edge energy of the output probability map\cite{r25}, which means the output probability map with the more misclassified area will have a large TV loss value of  $\mathcal{L}_{TV}$.  $\mathcal{L}_{TV}$ is defined as bellows: 

\begin{equation}
\mathcal{L}_{TV}(\hat{\mathbf{Y}}) =\sum_{i=1}^{H}\sum_{j=1}^{W}(\frac{(\hat{y}_{i,j}-\hat{y}_{i+1,j})^2}{H\times (W-1)} + \frac{(\hat{y}_{i,j}-\hat{y}_{i,j+1})^2}{(H-1)\times W})
\end{equation}

\section{Experiments}
\subsection{Dataset}

The dataset used in this paper is collected from 3D ultrasound scanning in the whole spine region. Then, ultrasound volume projection imaging (VPI) technique is utilized to generate 2-D mapped images. Ultrasound VPI images from 109 subjects with different degrees of scoliosis are used. The bone features are labeled by the medical experts. We randomly divided the dataset to a training set and a testing set with 80 and 29 samples respectively. The images are all in gray scales. The size of the ultrasound images corresponding to different patients is different. In the training set, all the images are resized to 250 $\times$ 1000 pixels. In the testing set, the original image size will be recorded and then input to the segmentation model after resized to 250 $\times$ 1000 pixels. Finally the segmentation mask will be resized to the same dimension as those of the original images. 

\subsection{Experimental Setups}

The proposed bone feature segmentation model was implemented using PyTorch\cite{r29}. In the training stage, random crop (that randomly selects the patch of the training images), horizontal flip (with 50\% probability to flip the input patch) as well as affine transformation were utilized to augment the training data to avoid overfitting. The network was trained for 50 epochs by the Adam optimizer \cite{r30} with a learning rate of 0.01 (weight decay = $1e^{-6}$).  All experiments were conducted by a system with one Nvidia GeForce RTX 2080Ti GPU.

\subsection{Results \& Discussion}

In this paper, we employed Dice score , Receiver Operating Characteristics (ROC) curve and Area Under the Curve (AUC) score to evaluate the segmentation performance. 

\subsubsection{Dice score} 
After obtained the segmentation results, we evaluate the proposed RSN-U-net by the bone feature area overlap of the segmentation results. Let $A_F$ and $A_G$ corresponding to the set of pixels of the bone feature segmentation result and the ground truth, respectively. We access the performance of our proposed method using Dice score\cite{r26},  which quantifies the area overlap between two regions, viz:

 \begin{equation}
 Dice(A_F, A_G) = 2\times\frac{A_F \cap A_G}{A_F + A_G}
 \end{equation}

\noindent A Dice score of 1 means the regions of $A_F$ and $A_G$ are identical. A Dice score of 0 means the regions of  $A_F$ and $A_G$ do not have any overlap.

\subsubsection{ROC curve and AUC score} 

We regarded the bone feature segmentation task as a pixel-level binary classification task. So the AUC-ROC score is utilized to measure the classification performance of our proposed segmentation model at various threshold settings. ROC is a curve indicating probability. AUC measures the degree of separability\cite{r27}. The ROC curve reflects the relationship between True Positive Rate (TPR) (y-axis) and the False Positive Rate (FPR) (x-axis). It also measures the discriminative capacity of a model. A higher AUC means a better model.

\begin{table}[t]
	\centering
	\renewcommand\arraystretch{1.2}
	\begin{tabular}{p{15mm}|p{15mm}p{10mm}}
		\hline
		Method& Dice Score&AUC \\
		\hline
		U-net& 0.7608&0.97 \\
		RSN-U-net & \textbf{0.7838}& \textbf{0.98}\\
		\hline
	\end{tabular}
	\caption{The comparison results with baseline U-net and the proposed RSN-U-net. }
	\label{t1}
\end{table}

\begin{figure}[tbp]
	\centerline{\includegraphics[keepaspectratio,width=0.38\textheight]{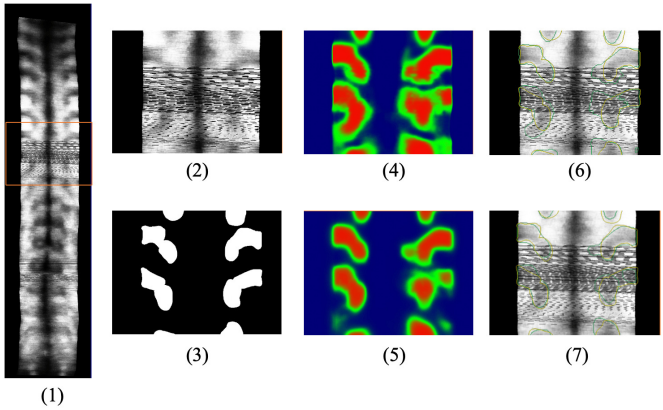}} 
	\caption{Visual comparison of the segmentation results on a part of ultrasound image with regular occlusion noise. (1) original VPI image; (2) ultrasound image with severe regular occlusion noise; (3) the ground truth mask; (4) the foreground probability map of the prediction result by U-net(visualized by heat map); (5) the foreground probability map of the prediction result by RSN-U-net; (6) segmentation result by U-net (Dice score: 0.7171); (7) segmentation result by RSN-U-net (Dice score: 0.7689) (Green line: segmentation result of the model; Yellow line: Ground Truth) (zoom in for better view).}
	\label{f4}
\end{figure}

Table \ref{t1} shows the Dice score and AUC score over 29 testing subjects of the U-net and the proposed RSN-U-net. 
Our proposed RSN-U-net achieve a 2.3\% average Dice score improvement as compared with the U-net model. The gap of the performance between U-net and RSN-U-net is more obvious in the case of input ultrasound image with speckle and regular occlusion noise. For the case that the input ultrasound image contains many speckles as shown in Fig. \ref{f3}, the RSN-U-net achieve a 3.17\% (= 0.8144 - 0.7827) Dice score improvement as compared with U-net. For the case that the input ultrasound image contains severe regular occlusion noise (Fig. \ref{f4}), our proposed RSN-U-net performs much better than U-net by a 5.18\% (= 0.7689 - 0.7171) Dice score improvement. The two special cases shows that our proposed RSN-U-net have a good robustness to speckle and regular occlusion noise. 

\begin{figure}[t]
	
	\centerline{\includegraphics[width=0.3\textheight, height=0.3\textwidth]{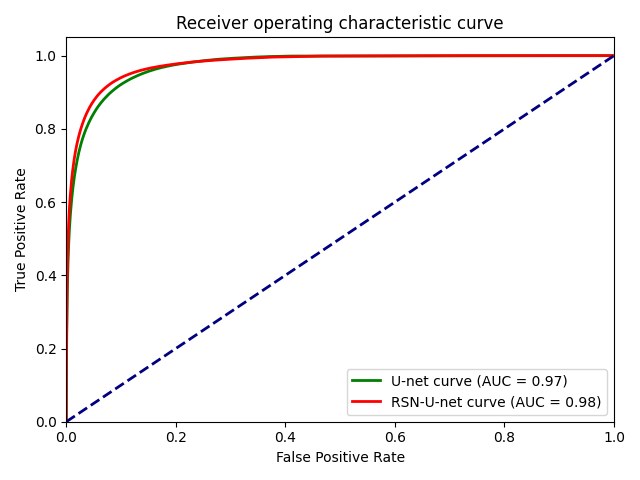}} 
	\caption{The ROC-AUC curve of the U-net and proposed RSN-U-net.}
	\label{f5}
\end{figure}

From the perspective of the classification, the proposed RSN-U-net also achieve 1\% AUC score improvement. As depicted in Fig. \ref{f5},  both testing methods perform very well according to the ROC curve. The ROC curve of the RSN-U-net is better than that of U-net, which means the classification ability of RSN-U-net outperforms that of U-net.


Fig. \ref{f3} shows a visualization of the bone feature segmentation results. From the probability maps, Fig. \ref{f3} (3) and (4), the proposed RSN-U-net shows low confidence in the non-bone region (the green and blue regions in the figure), while U-net tends to be confused by the speckled area and results in over-segmentations. From the segmentation results (Fig. \ref{f3} (5) and (6)), the proposed RSN-U-net performs better than the U-net.  Fig. \ref{f3} (7) shows a more detailed example of the segmentation results in a part of the input image with speckle. The U-net incorrectly classifies the speckled area as a bone area, while the RSN-U-net gives a correct segmental result of the bone feature.

Fig. \ref{f4} presents a detailed example of the segmentation results in a part of the input image with severe regular occlusion noise. According to the segmentation results (Fig. \ref{f4} (6) and (7)), the U-net fails to segment the bone feature. In contrast, the proposed RSN-U-net achieves a much better and more stable result even in the severe noisy areas. From the probability maps in Fig. \ref{f4} (4) and (5), we can see that owing to the introduction of the TV loss, the proposed RSN-U-net shows good robustness to high-frequency noise when compared with the U-net.

\section{Conclusion}

This paper propose a bone feature segmentation model for ultrasound spine image with high robustness to speckle and regular occlusion noise. After improving the loss function of the segmentation by introducing the total variance loss function, our proposed segmentation model (RSN-U-net) achieves a better performance (78.38\%) as compared with the baseline model (U-net). When using ultrasound images to measure the degree of scoliosis, it is very important to correctly segment the area of the bone. The ultrasound spine images tend to contain many speckles and regular occlusion noise which will increase the difficulty for experts to find out the bony region. Analysis of results shows that our model is robust to speckle and regular occlusion noise and can obtain a good segmentation result even the input is noisy.

In the future, we will invite experts to use the segmentation results of our model to measure the degree of scoliosis and compare it with the results obtained manually to further verify the reliability of our model. Further study would also focus on exploring the potential of the proposed model on setting up a fully automatic measurement system for accessing scoliosis.

\section{acknowledgment}

The work described in this paper was partially supported by Hong Kong Research Grant Council Research Impact Fund (R5017-18), and a grant from The Hong Kong Polytechnic University.

\bibliographystyle{IEEEtran}
\bibliography{ref_conf.bib}


\end{document}